\documentclass[12pt]{article}
\usepackage{epsf}
\usepackage{amsmath}
\usepackage{graphicx}
\usepackage{color}

\newcommand{\bmat}{\left(\begin{array}}
\newcommand{\emat}{\end{array}\right)}

\def\yzero{\smash{\hbox{$y\kern-4pt\raise1pt\hbox{${}^\circ$}$}}}

\def\beq{\begin{equation}}
\def\eeq{\end{equation}}
\def\beqa{\begin{eqnarray}}
\def\eeqa{\end{eqnarray}}

\def\th{\theta}

\def\-{\hphantom{-}}
\def\ov{\overline}
\def\s2{\frac{1}{\sqrt2}}

\def\beq{\begin{equation}}
\def\eeq{\end{equation}}
\def\beqa{\begin{eqnarray}}
\def\eeqa{\end{eqnarray}}

\def\IF{\relax{\rm I\kern-.18em F}}
\def\II{\relax{\rm I\kern-.18em I}}

\def\cn{{\cal N}}

\def\cam{{\cal M}}

\def\cf{{\cal F}}
\def\cd{{\cal D}}
\def\co{{\cal O}}

\def\Dsl{\,\raise.15ex\hbox{/}\mkern-13.5mu D} 

\def\IC{\bf C}
\def\IS{\bf S}

\def\NN{{\cal N}}
\def\CO{{\cal O}}

\def\k{\kappa}

\def\C{{\bf C}}

\def\G{\Gamma}

\newcommand{\drawsquare}[2]{\hbox{%
\rule{#2pt}{#1pt}\hskip-#2pt
\rule{#1pt}{#2pt}\hskip-#1pt
\rule[#1pt]{#1pt}{#2pt}}\rule[#1pt]{#2pt}{#2pt}\hskip-#2pt
\rule{#2pt}{#1pt}}

\newcommand{\fund}{\raisebox{-.5pt}{\drawsquare{6.5}{0.4}}}
\newcommand{\antifund}{\overline{\fund}}

\catcode`\@=11   
\newdimen\@rotdimen
\newbox\@rotbox  

\def\@vspec#1{\special{ps:#1}}
\def\@rotstart#1{\@vspec{gsave currentpoint currentpoint translate
   #1 neg exch neg exch translate}}
\def\@rotfinish{\@vspec{currentpoint grestore moveto}}
\def\@rotr#1{\@rotdimen=\ht#1\advance\@rotdimen by\dp#1%
   \hbox to\@rotdimen{\hskip\ht#1\vbox to\wd#1{\@rotstart{90 rotate}%
   \box#1\vss}\hss}\@rotfinish}
\def\@rotl#1{\@rotdimen=\ht#1\advance\@rotdimen by\dp#1%
   \hbox to\@rotdimen{\vbox to\wd#1{\vskip\wd#1\@rotstart{270 rotate}%
   \box#1\vss}\hss}\@rotfinish}%
\def\@rotu#1{\@rotdimen=\ht#1\advance\@rotdimen by\dp#1%
   \hbox to\wd#1{\hskip\wd#1\vbox to\@rotdimen{\vskip\@rotdimen
   \@rotstart{-1 dup scale}\box#1\vss}\hss}\@rotfinish}%
\def\@rotf#1{\hbox to\wd#1{\hskip\wd#1\@rotstart{-1 1 scale}%
   \box#1\hss}\@rotfinish}%
\def\rotate{\@ifnextchar[{\@rotate}{\@rotate[l]}}
\def\@rotate[#1]#2{\setbox\@rotbox=\hbox{#2}\@nameuse{@rot#1}\@rotbox}

\catcode`\@=12

\topmargin
-1.5cm
\textwidth
15.5cm
\textheight
23.5cm
\oddsidemargin
0.7cm
\evensidemargin
1.2cm

\begin{document}

\makeatletter
\@addtoreset{equation}{section}
\makeatother
\renewcommand{\theequation}{\thesection.\arabic{equation}}
\pagestyle{empty}
\rightline{ IFT-UAM/CSIC08-25}
\rightline{ CERN-PH-TH/2008-097}
\vspace{0.1cm}
\begin{center}
\LARGE{\bf Non-perturbative F-terms \\
 across lines of BPS stability  \\[12mm]}
\large{I. Garc\'{\i}a-Etxebarria$^{1,2}$, F. Marchesano$^{1}$, A.M. Uranga$^{1,2}$
\\[3mm]}
\footnotesize{{}$^1$ PH-TH Division, CERN 
CH-1211 Geneva 23, Switzerland\\
 
 \medskip
 
{}$^2$ Instituto de F\'{\i}sica Te\'orica UAM/CSIC,\\[-0.3em]
Universidad Aut\'onoma de Madrid C-XVI, 
Cantoblanco, 28049 Madrid, Spain \\[2mm] }
\small{\bf Abstract} \\[5mm]
\end{center}
\begin{center}
\begin{minipage}[h]{16.0cm}

We consider non-perturbative terms in the 4d effective action due to BPS D-brane instantons, and study their continuity properties in moduli space as instantons cross lines of BPS stability, potentially becoming non-BPS. We argue that BPS instantons contributing to the superpotential cannot become non-BPS anywhere in moduli space, since they cannot account for the required four goldstino fermion zero modes. At most they can reach lines of threshold stability, where they split into mutually BPS multi-instantons, as already discussed in the literature. On the other hand, instantons with additional fermion zero modes, contributing to multi-fermion F-terms, can indeed cross genuine lines of marginal stability, beyond which they lead to non-BPS systems. The non-BPS instanton generates an operator which is a D-term locally in moduli space, but not globally. This is due to a cohomological obstruction localized on the BPS locus, where the D-term must be written as an F-term, thus ensuring the continuity of the 4d contribution to the effective action. We also point out an interesting relation between lifting of fermion zero modes on instantons and 4d supersymmetry breaking.

\end{minipage}
\end{center}
\newpage
\setcounter{page}{1}
\pagestyle{plain}
\renewcommand{\thefootnote}{\arabic{footnote}}
\setcounter{footnote}{0}

\vspace*{1cm}

\section{Introduction}

An important aspect of the dynamics of string theory compactifications is the study of non-perturbative effects, arising from euclidean brane instantons (see e.g. \cite{Becker:1995kb, Witten:1996bn, Harvey:1999as, Witten:1999eg}). Indeed, recent applications of  D-brane instantons in type II compactifications (or F/M-theory duals) include, among others, mechanisms of moduli stabilization \cite{Kachru:2003aw,Denef:2004dm,Denef:2005mm}, generation of perturbatively forbidden couplings \cite{Blumenhagen:2006xt,Ibanez:2006da} (see also \cite{Haack:2006cy,Florea:2006si,Ibanez:2007rs}), generation of gauge field theory superpotentials \cite{Billo:2002hm,Akerblom:2006hx,Bianchi:2007fx,Argurio:2007vq,Bianchi:2007wy}, and realization of supersymmetry breaking \cite{Argurio:2007qk,Aharony:2007db,Aganagic:2007py}. Other aspects of D-brane instantons have been recently discussed in e.g. \cite{Billo:2007py,Akerblom:2007uc,bm,Blumenhagen:2008ji,Cvetic:2008ws,Matsuo:2008nu}. 

Some of these applications, like moduli stabilization, involve the use of the instanton-induced 4d operators throughout moduli space. The computation of such global expression for non-perturbative effects would seem feasible for 4d F-terms (e.g. superpotentials) which are holomorphic on all moduli and thus nicely behaved. However
this question is subtler than it seems, since the non-perturbative contributions to a 4d F-terms depends in principle on the spectrum of BPS instantons at each point in moduli space, and this spectrum can jump discontinuously across lines of BPS stability (see e.g. \cite{Denef:2001ix}), which are generically real codimension one loci in moduli space. It is therefore important to elucidate the behaviour of instanton effects across such loci, and the mechanisms to restore holomorphy of the 4d non-perturbative F-term.

This program was initiated in \cite{GarciaEtxebarria:2007zv} for instantons generating superpotential terms, and crossing lines of threshold stability,\footnote{We adopt the nomenclature in \cite{deBoer:2008fk} and distinguish between lines of threshold stability and of marginal stability. We refer to both of them as lines of BPS stability, in that the spectrum of BPS objects jumps at them.} namely real codimension one loci in moduli space, where a BPS brane splits into two (or more) mutually BPS branes, which beyond the line recombine back into another single BPS brane. Continuity of the superpotential requires novel kinds of contributions to the non-perturbative superpotential arising from multi-instanton effects. These multi-instanton effects have subsequently appeared in several new applications and contexts \cite{Ibanez:2007tu,Blumenhagen:2008ji,Cvetic:2008ws}.

Among the systems considered in \cite{GarciaEtxebarria:2007zv}, there is no example of BPS instantons contributing to the superpotential and crossing genuine lines of marginal stability, i.e. codimension one loci in moduli space, beyond which a BPS instanton turns into a (possibly multi-instanton) non-BPS system. In fact, such a situation is not possible, for a simple well-known reason. For an instanton to contribute to the superpotential, it must have exactly two fermion zero modes (to saturate the $d^2\theta$ superspace integration); while a non-BPS instanton breaks all supersymmetries and therefore has at least four goldstino fermion zero modes (which saturate the $d^4\theta$ superspace integration, so the instanton actually generates a 4d D-term). This simple observation, together with continuity of the non-perturbative effects, implies that it is not possible that a BPS instanton contributing to the superpotential is connected with a non-BPS instanton.\footnote{While this paper was in preparation, a revised version of \cite{Cvetic:2008ws} appeared, with results consistent with our statement.}

Equivalently, any BPS instanton which can cross a genuine line of marginal stability and become non-BPS cannot contribute to the non-perturbative superpotential. Such BPS instantons must have additional fermion zero modes, and therefore contribute to multi-fermion F-terms (denoted higher F-terms henceforth). The present paper is devoted to the study of the behaviour of non-perturbative higher F-terms across lines of marginal stability for the underlying instantons. Since higher F-terms are chiral operators and enjoy interesting holomorphy properties, they are expected to be well-behaved upon such crossings. 

This would seem to contradict the equally well-founded expectation that the non-BPS instanton generates a D-term. As we argue, there is no contradiction, for a simple but deep reason. As discussed in \cite{Beasley:2004ys,Beasley:2005iu}, higher F-terms are 4d operators which, when regarded as functions over moduli space, are associated to a non-trivial class of a certain cohomology, such that locally in moduli space can be written as integrals over all superspace, but not globally in moduli space. 

Thus the instanton amplitude is in such a cohomologically non-trivial class, when regarded globally in moduli space. Away from the BPS locus, the instanton amplitude is locally trivial in moduli space, and can be written as an integral over all superspace, a 4d D-term, in agreement with standard wisdom for non-BPS instantons. On the BPS locus, the amplitude reduces to  a chiral operator integrated over half superspace, an F-term, in agreement with standard wisdom for BPS instantons. In this sense, the cohomological obstruction to writing the amplitude as a globally defined D-term localizes on the BPS locus. Continuity and holomorphy of the 4d instanton amplitude are naturally described using these concepts.

We can also phrase in this language the fact that BPS instantons generating superpotentials cannot become non-BPS. A superpotential is non-trivial even locally in moduli space, in the sense that it cannot be written as an integral over all superspace. Therefore, a putative non-BPS instanton generating a superpotential would contradict standard wisdom of instanton physics, since such superpotential cannot be written as a D-term, even locally in moduli space.

The fact that only instantons with additional fermion zero modes can cross genuine lines of marginal stability (while instantons with two fermion zero modes have at most lines of threshold stability) leads to an interesting puzzle. Indeed, there are different mechanisms like, e.g. closed string fluxes, which can lift additional fermion zero modes and turn one kind of instanton into another. The resolution involves four-dimensional supersymmetry breaking in an interesting way. Consider a BPS instanton which can become non-BPS, and which therefore has additional fermion zero modes. Any mechanism which lifts the additional fermion zero modes and makes the instanton contribute to the superpotential (at least on the BPS locus), simultaneously triggers 4d supersymmetry breaking in the region of moduli space where the instanton is non-BPS. Since there is no supersymmetry in the background in this region, the instanton need not have four goldstino zero modes, thus avoiding a contradiction with the fact that the would-be goldstinos have been lifted. We present several realizations of this general argument, providing further support for our general picture.

\medskip

As we have mentioned, the prototypical example of line of marginal stability is provided by a BPS instanton, which splits into several BPS instantons, which subsequently misalign their BPS phases and define an overall non-BPS system. The discussion of how a multi-instanton  process reconstructs the amplitude of a single-instanton one is similar to  \cite{GarciaEtxebarria:2007zv}, and we will not pursue it further here. Indeed, most of the new conceptual issues are related to the instanton becoming a non-BPS system, rather than to its splitting. Hence, it is more illustrative to focus on simpler systems of instantons which are BPS on a real codimension one locus in moduli space, and are non-BPS away from it, with no splitting whatsoever. We regard this situation as another kind of line of BPS stability, with many features in common to lines of marginal stability. Indeed, as we discuss in one example, the resulting lessons, complemented with ideas in \cite{GarciaEtxebarria:2007zv}, suffice to describe lines of marginal stability with splitting of instantons.

The paper is organized as follows. In Section \ref{bw} we review some useful properties of higher F-terms. In Section \ref{isolated} we describe the example of an isolated $U(1)$ instanton, with a real codimension one BPS locus in moduli space, away from which it is non-BPS due to a misalignment of its calibrating phase. We describe the structure of its amplitude on the BPS locus, in the deep non-BPS regime, and their nice continuity in the near-BPS region. We show in Section \ref{lines} that these concepts carry over with little modification to lines of marginal stability involving splitting of instantons. In Section \ref{nfncbw} we discuss the instanton amplitude and a line of BPS stability in a system providing a D-brane realization of $N_f=N_c$ SQCD. In Section \ref{lifting} we present examples showing the correlation between fermion zero mode lifting and 4d supersymmetry breaking. Finally, Section \ref{conclu} contains some concluding remarks and outlook.

\section{Review of higher F-terms}
\label{bw}

In this section we briefly review some useful properties of multi-fermion F-terms (higher F-terms henceforth), following \cite{Beasley:2004ys,Beasley:2005iu}. 

BPS instantons with $2p$ additional fermion zero modes, beyond the two $\NN=1$ goldstinos, generate a multi-fermion F-term of the form
\beqa
\label{fterm}
\delta S \,&= &\, \int \! d^4 x \, d^2 \theta \; \omega_{\bar i_1 \cdots
\bar i_p \,\bar j_1 \cdots \bar j_p} \, (\Phi) \; \left(\bar D_{\dot{\alpha}_1} \mskip 2
mu\bar\Phi{}^{\bar i_1} \bar D^{\dot{\alpha}_1}\mskip 2 mu\bar\Phi{}^{\bar
j_1}\right) \cdots \left(\bar D_{\dot{\alpha}_p}\mskip 2 mu\bar\Phi{}^{\bar i_p} 
\bar D^{\dot{\alpha}_p}\mskip 2 mu\bar\Phi^{\bar j_p}\right)\,, \nonumber \\
& \equiv & \, \int \! d^4 x \, d^2 \theta \; \CO_\omega
\eeqa
where the field dependent tensor $\omega_{\bar i_1 \cdots \bar i_p
\,\bar j_1 \cdots \bar j_p}$ is antisymmetric in the $\bar i_k$ and also in the $\bar j_k$, and symmetric under their exchange. Formally it can be regarded as a section of $\bar\Omega^p_{\cal M} \otimes \bar\Omega^p_{\cal M}$.

The conditions that $\delta S$ is supersymmetric and a non-trivial
F-term implies that $\omega$ belongs to a non-trivial cohomology class
in moduli space, for the non-standard cohomology defined below. The
condition that $\delta S$ is supersymmetric is that $\CO_\omega$ is
chiral, namely annihilated by the supercharges ${\ov
  Q}_{\dot{\alpha}}$. This requires that $\omega$ is holomorphic,
namely closed under ${\ov \partial}$. On other hand, even if $\delta
S$ is supersymmetric, it may represent a trivial $F$-term. Though
written in (\ref{fterm}) in the form $\int d^2\theta(\dots)$, it may
be that $\delta S$ can be alternatively written $\int
d^4\theta(\dots)$, in other words as a $D$-term. In fact, if
$\CO_\omega=\{\bar Q{}_{\dot\alpha},[\bar Q{}^{\dot \alpha},V]\}$ for
some $V$, then $\CO$ is trivially chiral and one can write $\delta
S=\int d^4x\,d^4\theta\, V$. One must therefore impose an equivalence
relation on the space of operators $\CO_\omega$, under which it is
considered trivial if $\delta S$ is equivalent to a $D$-term. The
equivalence relation is of the form \beqa \omega_{\bar i_1 \cdots \bar
  i_p \, \bar j_1 \cdots \bar j_p} \,\sim\, \omega_{\bar i_1 \cdots
  \bar i_p \, \bar j_1 \cdots \bar j_p} \,+\, \nabla_{[{\bar i_1}}
\xi_{\bar i_2 \cdots {\bar i_p}] \, \bar j_1 \cdots \bar j_p} \,+\,
\left(\bar i_k \leftrightarrow \bar j_k\right)\,.
\label{equiv}
\eeqa with symmetrization of the $i$ and $j$ indices in the term
involving $\xi$. This condition is actually somewhat stronger in
principle than being a D-term, and just requires that $\delta S$
cannot be written as an integral over 3/4 of superspace.

As a simple example, consider the familiar statement that any
correction $\delta K$ to the K\"ahler form can be trivially rewritten
as an $F$-term correction upon performing half the integral over
superspace: \beqa \int \! d^4 x \, d^4 \theta \; \delta K = \int \!
d^4 x \, d^2 \theta \; \bar D^2 \! \delta K\, = \int \!  d^4 x \, d^2
\theta \; \nabla_{\bar i} \nabla_{\bar j} \delta K \, \left(\bar
  D\mskip 2 mu\bar\Phi{}^{\bar i} \cdot \bar D\mskip 2
  mu\bar\Phi{}^{\bar j}\right) \eeqa where $\nabla$ is the covariant
derivative in field space. Hence 4-fermion F-terms with $\omega_{\bar
  i\, \bar j} = \nabla_{\bar i} \nabla_{\bar j} \delta K$ are trivial
and actually correspond to D-terms.

The condition that $\omega$ defines a supersymmetric operator (i.e. closed under ${\ov\partial}$) but a non-trivial F-term (i.e. not exact in the above sense) implies that it defines a non-trivial class with respect to a cohomology. Because of the symmetrization in (\ref{equiv}), the cohomology does not correspond to the usual Dolbeault cohomology.
Still, the only information we need to keep in mind for our purposes is that
the cohomology groups associated to higher F-terms are locally trivial in moduli space. Namely, it is always possible to write the interaction as a D-term locally in moduli space, but there may be a cohomological obstruction to doing it globally in moduli space. The interaction must therefore be written as an F-term of the form (\ref{fterm}), modulo pure D-terms.

\section{The isolated $U(1)$ instanton}
\label{isolated}

In this section we consider the simplest system of an instanton which can become non-BPS, by simple misalignment of its BPS phase. As argued in the introduction, it cannot contribute to the superpotential but rather to higher F-terms of the kind studied in previous section. This example of line of BPS stability is simple in that the instanton does not split, but still leads to several important lessons, which apply with little modification to more involved lines of marginal stability where instantons split,  as we show in section \ref{lines}.

\subsection{The setup}

Let us consider a Calabi-Yau compactification with an orientifold
projection, and a D-brane instanton which is not mapped to itself
under it (we denote the instanton $U(1)$ hereafter). We consider the
instanton to be described by an A- or B- brane in type IIA or type IIB
respectively, so that the holomorphic part of the supersymmetry
conditions are satisfied. Such instanton has a real codimension one BPS locus in
moduli space (complex structure in IIA, Kahler in IIB), where its BPS
phase aligns with the $\NN=1$ subalgebra preserved by the orientifold
plane. At this locus the instanton breaks half of the supersymmetries
of the 4d $\NN=1$ background, while away from it it breaks all
supersymmetries. Finally, for simplicity we consider the instanton to
be rigid, and isolated, i.e. not intersecting its orientifold image,
so that the zero mode structure reduces to the universal sector of
four translational bosonic zero modes and four fermionic zero modes.

It is easy to devise simple realizations using euclidean D2-branes wrapped on 3-cycles in a IIA orientifold compactification, and we phrase the discussion in these terms (although it is straightforward to do it for general situations in IIB). For a fully explicit realization, we can use the geometries described in appendix \ref{oovafa}, see Figure \ref{misalign}.

\begin{figure}
\epsfysize=2.8cm
\begin{center}
\leavevmode
\epsffile{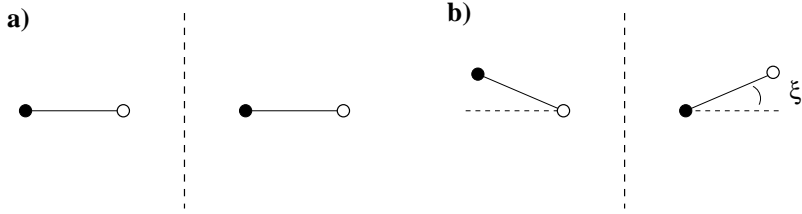}
\end{center}
\caption{\small A rigid isolated $U(1)$ instanton (a) on the BPS locus (with respect to the $\NN=1$ supersymmetry preferred by the orientifold plane, shown as a dashed line), and (b) when non-BPS due to a misalignment of its BPS phase. As usual, black and white dots denote the degenerations of the double $\IC^*$ fibration in the geometries discussed in Appendix \ref{oovafa}.}
\label{misalign}
\end{figure}

Let us fix some notation for the relevant closed string moduli. Consider a D2-brane wrapped on a 3-cycle $A$ and its orientifold image wrapped on $A'$. We also introduce the dual cycles $B$ and $B'$. Before the orientifold projection, the relevant closed string sector is given by two hypermultiplets, with 4d bosonic components
\beqa
t\, =\, \int_A {\rm Re\,}\Omega\,+\, i\, \int_A C_3 \quad , \quad u \, =\,  \int_B {\rm Im\,}\Omega\,+\, i\, \int_B C_3 
\eeqa
and similar primed fields for the primed cycles. In 4d $\NN=1$ terms we have the chiral multiplets  
\beqa
T\,=\, t\,+\, \theta\, \psi \quad , \quad \Sigma\, =\, u\,+\,\theta \, \chi
\eeqa
and similar primed fields. The combinations invariant under the orientifold are $T+T'$ and $\Sigma-\Sigma'$, which we also denote $T$, $\Sigma$ by abuse of language.
For future convenience, we also define
\beqa
\xi\, =\, {\rm Re}\, u\, =\, \int_B {\rm Re\,}\Omega
\eeqa

\subsection{On the BPS locus}
\label{sec:U(1)-bps}

Let us consider the BPS locus, where the instanton is BPS with respect
to the $\NN=1$ preferred by the orientifold plane, see Figure
\ref{misalign}a. As mentioned, there are only the four zero modes,
which we denote $\theta$, ${\ov \tau}$, providing the goldstinos of
the $\NN=2$ susy in the absence of the orientifold plane. The modes
$\theta$ are goldstinos of the $\NN=1$ supersymmetry of the
compactification, and are saturated by the $d^2\theta$ integration
over half of superspace of the non-perturbative 4d F-term expected
from a BPS instanton. The modes ${\ov \tau}$ are additional fermion
zero modes, would-be goldstinos of an orthogonal $\cn = 1'$ broken by
the orientifold plane. Their presence implies that the F-term is not a
superpotential, but rather a higher F-term.

\begin{table}[ht]
\centering
\begin{tabular}{|c|c|}
\hline
${\cal N}=1$  & ${\cal N}=1'$   \\
\hline \hline
$    \theta^{\alpha}  $ & $ \tau^{\alpha}  $   \\ \hline
$ {  \ov \theta^{\dot \alpha} } $ & $ {\ov \tau^{\dot \alpha} } $   \\
\hline
\end{tabular}
\caption{\small Universal fermionic zero modes $\theta^{\alpha}, \ov \tau^{\dot \alpha}$  of an instanton associated with the breaking of an $\{\cn = 2\} = \{\cn= 1\} \oplus \{\cn = 1'\}$ SUSY algebra into the ${\cal N}=1$ subalgebra preserved by the orientifold. For an anti-instanton the universal zero modes correspond to ($\tau^{\alpha}, \ov \theta^{\dot \alpha})$.
\label{zero} } 
\end{table}

Higher F-terms from D-brane instantons have been considered e.g. in \cite{Blumenhagen:2007bn}. Let us nevertheless offer a heuristic derivation of the structure of the amplitude, well suited for our purposes. Since it is necessary to integrate over the instanton fermion zero modes, one needs to write the instanton action with its full dependence on $\theta$, ${\ov\tau}$. Namely, one needs to include the supersymmetric variations of the instanton action with respect to the corresponding supersymmetries. We thus have
\beqa
S_{\rm inst}\, =\, t\, +\, \theta\, \delta_\theta t\, +\, {\ov \tau}\, {\ov \delta}_{\ov\tau}\, t \,+ \,( \theta \,\delta_{\theta} )\,  ({\ov\tau} \, {\ov\delta}_{\ov\tau})\, t
\eeqa
where the $\delta$'s denote the variations with respect to the $\NN=2$ supersymmetries. Namely
\beqa
\delta_\theta\, t\, =\, \psi \quad, \quad {\ov \delta}_{\ov\tau}\, t \, =\, {\ov\chi} \quad ,\quad
\delta_{\theta^\alpha} {\ov\delta}_{\ov\tau^{\dot{\alpha}}}\,t\, =\, (\ov\sigma^\mu)_{\alpha\dot{\alpha}} \, \partial_\mu \, u
\eeqa
where the last expression follows from the observation that ${\ov \chi}$ is the lowest component of the chiral multiplet
\beqa
{\ov D}_{\dot{\alpha}} {\ov \Sigma}={\ov\chi}_{\dot{\alpha}} + (\theta{\ov \sigma}^\mu)_{\dot{\alpha}} \, \partial_\mu u
\eeqa
Notice that the value of the parameter $\xi$ does not appear in this action (only its derivatives do).\footnote{Incidentally, one can also heuristically understand the derivative coupling $u$, or rather its imaginary part, as follows. The field ${\rm Im}\, u$ is related to the integral of RR potentials over the wrapped cycle, so $\theta{\ov\sigma}^\mu{\ov\tau} \partial_\mu u$ comes from the coupling of fermionic fields to RR field strength in the D-brane fermionic action (see e.g. \cite{Martucci:2005rb}).} Indeed, it controls the misalignment with respect to the $\NN=1$ supersymmetry, so it is implicitly set to zero in the above expression in order to keep our instanton BPS. Note also that we are already implicitly using supermultiplets under the $\NN=1$ supersymmetry preferred by the orientifold, and that the above discussion can be carried out similarly for the orientifold image. Including the orientifold and its image, the complete instanton amplitude reads
\beqa
\int\, d^2\theta\, d^2{\ov\tau} \, e^{-(\,T\,+\, {\ov\tau}\, {\ov D}{\ov \Sigma}\,)}\,=\, \int\,d^2\theta\, e^{-T}\, {\ov D}{\ov \Sigma}\cdot {\ov D}{\ov \Sigma}
\label{fisolated}
\eeqa
This is a supersymmetric higher F-term, due to holomorphy of the coefficient. Its non-triviality is encoded in the fact that the instanton is BPS and has no fermion zero modes along the ${\ov \theta}$ direction in superspace.

\subsection{Away from the BPS locus}

Let us move the system away from the BPS locus by turning on the real closed string modulus ${\rm Re}\, u$ in the multiplet $\Sigma$, which in the IIA realization controls the size of the 3-cycle B. As is manifest in Figure \ref{misalign}b, where it corresponds to moving the white degeneration away from the real axis, this modulus controls the misalignment of the BPS phase of the instanton.

In general, such misaligned instantons have a very non-holomorphic structure. For instance, the bosonic part of their classical action (the wrapped volume of the instanton and its orientifold image) is a non-holomorphic function of the moduli of the $\NN=1$ theory. To be more specific, consider the IIA setup, with a D2-brane instanton wrapped on a 3-cycle calibrated with respect to $e^{i\xi}\Omega$, where $\xi$ is the phase of the central charge.\footnote{Abusing language, we use the same notation for the BPS phase and for ${\rm Re}\, u$, even though the are the same only in the near BPS regime of next section.} The real part of the classical action of such instanton is given by ${\rm Re}\,(e^{i\xi} \Omega)|_{\Pi}$. This can be phrased in terms of its classical action on the BPS locus, ${\rm Re}\, \Omega|_\Pi$, which is a holomorphic function of the moduli, by using the calibration condition
\beqa
{\rm Im} \, (e^{i\xi} \,\Omega)|_\Pi\, =\, 0 \quad \longrightarrow \quad {\rm Im}\, \Omega|_\Pi\, =\, -\tan\xi\, {\rm Re}\, \Omega|_\Pi
\eeqa
so that
\beqa
{\rm Re}\, (e^{i\xi}\,  \Omega)|_\Pi\, =\, \cos\xi\, {\rm Re}\, \Omega|_\Pi\, -\sin \xi\,{\rm Im}\, \Omega|_\Pi\, =\frac{{\rm Re}\, \Omega|_\Pi}{\cos \xi}
\eeqa
which is a non-holomorphic function of the moduli, since $\xi$ is real.

A second general feature of these non-BPS instantons is that they have
at least four exact fermion zero modes, the goldstinos generated by
acting on the instanton with the four supercharges it breaks. Hence,
in agreement with standard wisdom, such non-BPS instantons generate
D-terms. Indeed, we can be more explicit in this respect, by
considering the structure of fermion zero modes for a non-BPS instanton with a
non-zero, constant phase $\xi$. The instanton is supersymmetric with respect
to the $\NN=1$ subalgebra of the underlying $\NN=2$ defined by the
phase $\xi$. The fermion zero modes, denoted $\theta'$ and ${\ov
  \tau}'$, are associated to the orthogonal supercharges. This primed
modes can be rewritten in terms of the unprimed modes of the reference
$\NN=1$ subalgebra by performing a spinorial rotation of angle
$\xi$:
\beqa
{\ov \tau}' & = & \cos (\xi/2) \, {\ov\tau}\, +\, \sin(\xi/2)\,{\ov
  \theta} \notag\\
\theta' & = & \cos (\xi/2) \, \theta\, +\, \sin(\xi/2)\,\tau
\label{zerorot}
\eeqa
This expression can be understood easily by noticing that the
preserved goldstinos for a rotated brane are solutions of the
equation \cite{Berkooz:1996km}:
\beqa
\varepsilon_L = R\Gamma_{Dp}R^{-1} \varepsilon_R,
\label{gammarot}
\eeqa
where $\G_{Dp}$ is a product of $\G$-matrices along the worldvolume of 
the unrotated brane and $R$ is the rotation matrix
relating the rotated system to the unrotated one. This equation
implies that the susys preserved by the rotated instanton are simply a
rotation of the solutions for the aligned instanton, as in (\ref{zerorot}).

In the computation of the instanton amplitude, the fermion zero mode
integration measure $d^2\theta'\, d^2{\ov\tau}'$ can be expressed
in terms of variables natural in the 4d $\NN=1$ supersymmetry of the
compactification, by using\footnote{An analogue for bosonic variables
  is to consider a 2-plane parameterized by $x,y$, and introduce a
  system rotated by an angle $\xi/2$, namely $x' = \cos (\xi/2) \,
  x\,+\, \sin(\xi/2)\, y$, $y' = -\sin(\xi/2)\, x\,+\, \cos(\xi/2)
  y$. An integral along the $x'$ axis (hence $y'=0$ and so $y=
  \tan(\xi/2)x$) can be traded to an integral over $y$ using
  $dx'=dy/\sin(\xi/2)$.} $d^2{\ov \tau}'= \sin^2(\xi/2) d^2{\ov
  \theta}$ (and similarly $d^2\theta'=\cos^2(\xi/2)\, d^2\theta$).
Thus for non-zero BPS phase, the instanton generate terms which can be
written as integrals over all of superspace $\int d^2\theta
d^2{\ov\theta}\, (...)$, so the non-BPS instanton generates a D-term.

However, regarding the instanton induced 4d operator as a global
function over moduli space, the above change of variables signals a
pathology at the BPS locus, where $\xi=0$. The goldstino ${\ov\tau}$
has zero component along ${\ov\theta}$, and therefore the instanton
effect cannot be written as a D-term on this locus. In the language of
section \ref{bw}, the D-term produced by the instanton is in a
non-trivial class of the Beasley-Witten cohomology, with the
cohomological obstruction localized on the BPS locus, where the
contribution must be written as a genuine F-term.

This observation, to be developed in the next paragraph, underlies the
continuity of the non-perturbative contribution across the BPS
line. The generation of the D-term away from the BPS locus, as
dictated by physics of non-supersymmetric instantons, does not
contradict the generation of an F-term at the BPS locus, as dictated
by physics of supersymmetric instantons.

\subsection{The near BPS regime}

In order to show that the D-term generated by the non-BPS instanton
reduces to the required F-term at the BPS locus, we consider the
instanton slightly away from the BPS locus. Treating the amplitude in
an expansion in $\xi$, the ${\cal O}(\xi^0)$ piece reduces to the
genuine F-term of the BPS locus. Higher order contributions should
correspond to globally defined D-terms.

Consider moving slightly away from the BPS locus by moving in closed string moduli space in the direction $\xi$. From the viewpoint of the instanton,  this corresponds, in first approximation, to turning on a world-volume Fayet-Iliopoulos term. Thus at leading order the interesting couplings on the instanton world-volume theory are
\beqa
\xi\, D\quad ; \quad {\ov \chi} {\ov\tau} 
\label{ficoupling}
\eeqa
These can be understood by analogy with the terms obtained from coupling a field-dependent FI term in 4d $\int d^4\theta (\Sigma+{\ov \Sigma}) V$. The first term, once the auxiliary field $D$ is integrated out gives rise to an ${\cal O}(\xi^2)$ term, corresponding to the quadratic term in the expansion of the non-BPS instanton classical action ${\rm Re}\, \Omega|_\Pi/\cos\xi$. We drop it in our ${\cal O}(\xi^0)$ computation.
Notice also that the fermion zero modes $\theta'$, ${\ov\tau}$, reduce to $\theta$, ${\ov\tau}$ at this order.

The full instanton action, completing the $\theta$-dependence on the second term in 
(\ref{ficoupling}) (i.e. promoting ${\ov\chi}$ to the 4d supermultiplet ${\ov D}{\ov\Sigma}$) is
\beqa
S_{\rm inst}\, =\,T\,+\,  {\ov \tau}\, {\ov D}{\ov \Sigma}
\label{FIferm2}
\eeqa
thus reproducing the contribution on the BPS locus, which arises as a manifest F-term. 
\beqa
 \int d^4x\,d^2\theta\, d^2{\ov\tau} \, e^{-(\,T\, +\, {\ov\tau}\,{\ov D}{\ov \Sigma}\,)}\, =\,  \int d^4x\,d^2\theta \, e^{-T} \, {\ov D}{\ov \Sigma}\,{\ov D}{\ov \Sigma}
\eeqa
If preferred, in the above analysis we could have kept the integration over ${\ov\theta}$ instead of ${\ov\tau}$, in order to make the connection with the situation far away from the BPS locus. The result is the same, simply noticing the cancellations between $\xi$-dependent factors (to the corresponding order) between the fermion zero mode measure, and the insertions required to saturate ${\ov\theta}$. 

In conclusion, the complete instanton amplitude reduces to the above F-term on the BPS locus. Thus, the complete instanton amplitude and the F-term on the BPS locus correspond to F-terms in the same Beasley-Witten cohomology class. This implies that the higher order terms in $\xi$ correspond to pure D-terms, which are globally defined over moduli space, and that the genuine F-term contribution of the instanton is given by the above expression.

\subsection{The global picture}

Let us recap and refine the global picture. The instanton generates an
F-term contribution in a cohomologically non-trivial class of the
Beasley-Witten cohomology. This F-term is trivial away from the BPS
locus and can be expressed as an integral over all superspace in
agreement with the standard wisdom that non-BPS instantons generate
non-perturbative D-terms. The instanton amplitude on the BPS locus is
given by (\ref{fisolated}), and defines a genuine F-term, in agreement
with the standard wisdom for BPS instantons.

Let us comment on a possible source of confusion, regarding the
holomorphy of the instanton induced F-term as a function of moduli, in
particular of moduli which take the instantons away from their BPS
locus. Clearly the complete instanton amplitude does not depend
holomorphically on the 4d moduli, in particular on the moduli which
take the instantons away from their BPS locus. On the other hand, the
BPS amplitude (\ref{fisolated}) does have a nice holomorphic
structure, but, extending it trivially to the complete moduli space,
is not the complete expression for the instanton amplitude away from
the BPS locus. The crucial point is that both expressions are two
representatives of the same cohomology class, and thus differ by a
globally defined D-term. Therefore, all the non-holomorphies in the
complete instanton amplitude can be assigned to the part that
corresponds to a globally defined D-term. The genuine F-term is
therefore holomorphic in all moduli, and in particular
(\ref{fisolated}) provides a particular representative which makes
this property manifest. It is in this sense that holomorphy of the
non-perturbative F-term from the instanton is preserved across the
line of BPS stability. Although we phrased the discussion in terms of
our particularly simple example, the argument applies in full
generality to any other situations, including instantons that split at
lines of marginal stability, as we argue in the next section.

\subsection{Application to lines of marginal stability}
\label{lines}

In order to show that the lesson learned in the previous section applies also to instantons that split at lines of marginal stability, we discuss the simplest situation of this kind. Using a type IIA language, consider a system of two rigid $U(1)$ instantons, wrapped on two cycles $A_1$, $A_2$, with one chiral intersection, and away from the orientifold planes so that they do not intersect the orientifold image of the system. 

Such systems are easy to engineer in toroidal orientifold examples. Also, although the geometries in appendix \ref{oovafa} realize naturally only non-chiral intersections, it is straightforward to construct examples with such chiral intersections in closely related geometries. Consider a complex plane $z$ over which we fiber a $\C^*$ degenerating at points labeled $z=a_i$, times an elliptic fiber, with $(p_i,q_i)$ 1-cycles degenerating at points labeled $z=b_i$. Our system of interest can be realized by considering a geometry with one $a$-type degeneration and two $b$-type degenerations, with $(p_1,q_1)=(1,0)$, $(p_2,q_2)=(0,1)$. The two relevant 3-cycles $C_1$, $C_2$ are obtained by considering segments on the $z$-plane joining $[a,b_1]$, and $[a,b_2]$, respectively, and 
over which we fiber the circle in $\IC^*$ fiber and the 1-cycles $(1,0)$ and $(0,1)$ on the elliptic fiber, respectively. The 3-cycles have intersection numbers $[C_1]\cdot [C_2]=1$. By locating the degenerations in suitable locations on the $z$-plane the systems can be taken to be BPS, or misaligned by an arbitrary amount. It is also straightforward to add orientifold planes and other required ingredients. Incidentally, the above double fibrations have appeared as the mirrors of systems of D-branes at singularities, see e.g. \cite{Hanany:2001py,Feng:2005gw}. In this language, the above realization of our system
can be regarded as considering two fractional branes on the complex cone over $dP_1$.
In any event, we can proceed with our discussion in full generality, independently of the details of specific realizations.

The zero mode content of the instanton system is given by the two universal sets $x_i^\mu$, $\theta_i$, ${\ov\tau_i}$, for $i=1,2$, and bosonic and fermionic zero modes at the instanton intersection $m$, $\psi$, and ${\ov m}$, ${\ov\psi}$, with charges $(+1,-1)$ and $(-1,+1)$ under the two $U(1)$'s.

The system has a line of marginal stability, which is controlled by a real parameter. For 
simplicity we may keep the 3-cycle $A_2$ aligned, then the relevant real parameter is
$\xi\, =\, \int_{B_1}\, {\rm Im}\, \Omega$, with an additional term $-\int_{B_1'} {\rm Im}\, \Omega$ if we include the orientifold image system.\footnote{In more involved situations with two parameters $\xi_1$, $\xi_2$, the same role is played by  $\xi=\xi_1-\xi_2$.}
The line of marginal stability is located at $\xi=0$. We focus on its neighbourhood, namely small $\xi$, where its effect can be described as a world-volume FI term for $U(1)_1$. The discussion of the world-volume action in this regime can be carried out similar to \cite{GarciaEtxebarria:2007zv}. We have the couplings 
\begin{equation}
S_{\rm 2-inst}\, =\, |x_1-x_2|^2\, |m|^2\, +\, i (x_1^\mu-x_2^\mu)\, {\ov \psi} \,\sigma_\mu\, \psi\, +\,  \psi\,  (\theta_1-\theta_2)\, {\ov m} \, +\, {\ov \psi}\, ({\ov \tau}_1-{\ov\tau}_2) \,m
\end{equation}
In addition, there are couplings involving fields in the FI multiplet $\Sigma$
\beqa
V_D\, =\, (\, m{\ov m} \, -\, \xi\,)^2 \quad ; \quad S_{\ov\tau_1}\, =\, {\ov\tau}_1\, {\ov D}{\ov\Sigma}
\eeqa

Consider the system for $\xi>0$ which corresponds to the BPS side. Heuristically, the bosons $m, {\ov m}$ acquire a vev to minimize the scalar potential $V_D$. The instantons recombine into a single rigid BPS $U(1)$ instanton. The vev accordingly also freezes the two instantons at the same position $x_1=x_2$ in space, and in the fermionic coordinates $\theta_1=\theta_2$, $\tau_1=\tau_2$, by making massive their difference fields (along with $\psi$, ${\ov \psi}$). We are left with a recombined BPS instanton with a universal sector of translational zero modes $x_1-x_2$, and fermion zero modes $\theta=\theta_1+\theta_2$, ${\ov \tau}={\ov\tau}_1+{\ov \tau}_2$. The latter pull out two insertions of ${\ov D}{\ov\Sigma}$, so that, denoting $T=T_1+T_2$, the instanton amplitude is
\beqa
\int\, d^4x\, d^2\theta \, e^{-T}\, {\ov D}{\ov\Sigma}\, {\ov D}{\ov\Sigma}
\label{fmarginal}
\eeqa
as expected for the recombined BPS rigid instanton (notice that $\Sigma$ can be regarded as controlling the 3-cycle dual to $A=A_1+A_2$). In a more proper treatment, $m$, ${\ov m}$ are not frozen at their vevs, but rather one integrates over these bosonic modes. The above heuristic computation, however, provides the saddle point approximation to the instanton computation.

At the BPS locus $\xi=0$, we have a system of two mutually BPS
instantons, and the amplitude (\ref{fmarginal}) is reconstructed by a
2-instanton process involving both. In this computation, the above
saddle point approximation is not valid (since it would fix $m=0$ and
all fermion interactions would disappear), but the correct computation
by integrating over bosonic zero modes allows to keep the interactions
and saturate all fermion zero modes just as in the above
discussion. Hence, one recovers the same F-term structure
(\ref{fmarginal}).

Finally, for $\xi<0$ it is not possible to cancel the world-volume
D-term scalar potential, and we have a system of two instantons with
different BPS phases, thus defining an overall non-BPS system. The
non-cancellation of $V_D$ reflects the non-holomorphic dependence of
the total wrapped volume on the moduli. Nevertheless, the discussion
of fermion zero mode saturation is exactly as in the previous single
instanton situation. We are left with a non-BPS instanton, with
fermion zero modes $\theta=\theta_1+\theta_2+{\cal O}(\xi)$ and
${\ov\tau}={\ov\tau}_1+{\ov \tau}_2+\sin(\xi/2) {\ov\theta}$. The fact
that ${\ov \tau}$ picks up a component along ${\ov\theta}$ allows to
write the instanton amplitude as a D-term, as expected for a non-BPS
instanton, with no discontinuity at the BPS locus, since the non-BPS
instanton amplitude reproduces the F-term (\ref{fmarginal}) from its
${\cal O}(\xi^0)$ contribution.

As announced, the main concepts involved in the study of instantons at lines of marginal stability can be already obtained from simpler systems with lines of BPS stability where instantons do not split. We therefore focus on the latter kind of systems to continue discussing other new features, with the understanding that their discussion in genuine lines of marginal stability is possible, and very similar.

\section{The $N_f=N_c$ SQCD instanton from branes}
\label{nfncbw}

In the above discussion, the zero modes ${\ov \tau}$ played a
fundamental role, in that they allowed an interpolation between
fermion zero modes with the interpretation of goldstinos ${\ov
  \theta}$ (away from the BPS locus) and extra fermion zero modes
making the non-perturbative contribution a higher F-term rather than a
superpotential (at the BPS locus). It would seem crucial that the
${\ov \tau}$ are unlifted, and that the argument does not apply to
instantons where these models are lifted by interactions. In this
section we argue that in fact the discussion continues to be valid in
such cases. We describe this in an illustrative example, the D-brane
realization of the $N_c=N_f$ SQCD instanton. Since this example
provides a stringy\footnote{For a recent detailed study of the
  instanton in $N_f=N_c$ SQCD in a string theory realization, see
  \cite{Matsuo:2008nu}.} realization of the well-studied system in
\cite{Beasley:2004ys} (modulo the gauging of the center of mass $U(1)$
in our case), our discussion is sketchy and targeted to illustrating
the physics of the modes ${\ov\tau}$. In fact we carry out most
computations in the case $N_f=N_c=1$, where there is no classical
field theory interpretation for the instanton (it is stringy in the
sense of \cite{Petersson:2007sc}), since the latter is not essential
to our purpose.

\subsection{The setup}

\begin{figure}
\epsfysize=2.5cm
\begin{center}
\leavevmode
\epsffile{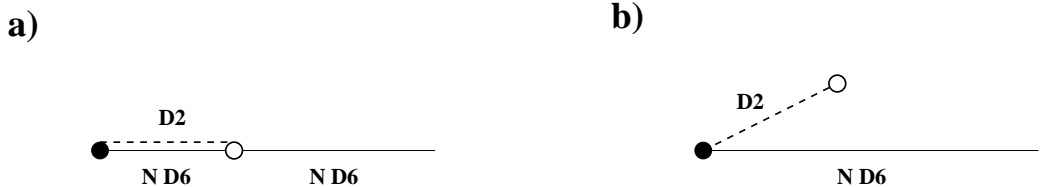}
\end{center}
\caption{\small D-brane realization of the $N_f=N_c$ SQCD theory in a
  geometry of the kind in Appendix \ref{oovafa}. The construction
  shows that instanton on its BPS locus (a) and away from it
  (b). Notice that gauge D-branes recombine and remain BPS through the
  process.}
\label{nfnc}
\end{figure}

Consider a D-brane instanton with the interpretation of a gauge
instanton in an $N_f=N_c$ SQCD theory. Namely, in the IIA setup we
consider $N_c=N$ D6-branes wrapped on an rigid 3-cycle $A$, and
$N_f=N$ D6-branes wrapped on a (possibly non-compact) 3-cycle $A_{\rm
  flav.}$, with a non-chiral intersection with $A$. The instanton of
interest is a D2-brane wrapped on $A$. The setup is easy to engineer
using the geometries described in Appendix \ref{oovafa}, by wrapping
sets of $N$ D6-branes on a compact and a non-compact 3-cycles as shown
in Figure \ref{nfnc}.\footnote{In this case the gauge D-branes suffice
  to define a preferred $\NN=1$ supersymmetry, so the presence of the
  orientifold plane is not necessary. If desired for any other reason,
  an orientifold plane could be added away from the D-brane system,
  along with the corresponding orientifold images.} As in previous
examples, there is a real parameter, given by the size of the 3-cycle
$B$ dual to $A$, which controls the misalignment of the instanton, see
Figure \ref{nfnc}b. Note that the misalignment of the instanton does
not imply a misalignment of the gauge D-branes, since the latter can
recombine and remain BPS throughout moduli space. This recombination
is important, as we will recall in Section~\ref{nfncnear}.

\subsection{At the BPS locus}

\subsubsection*{The Beasley-Witten higher F-term for $N_c=N_f=1$}

Let us sketch the fermion zero mode structure and their saturation in
the BPS case. In addition to the four universal fermion zero modes
$\theta$, ${\ov\tau}$, there are bosonic and fermionic zero modes
charged under the instanton world-volume $U(1)$.  We have fermion zero
modes $\beta$, ${\ov \beta}$ from the instanton to the color 4d
spacefilling D-branes, and $\eta$, $\ov \eta$ from the instanton to
the flavor D-branes. In addition we have bosonic zero modes from the
instanton to the color D-branes. Our description of the bosonic zero
modes will be sketchy, and we simply introduce the relevant ones for
fermion mode saturation which we denote by $m$, ${\ov m}$.  The
charges of these fields under the different groups are {\small
\begin{center}
\begin{tabular}{|c||c|c|c|}
\hline
Field & $U(1)_{\rm inst}$ & $U(N_c)$ & $U(N_f)$ \cr
\hline
\hline
$m$, $\beta$ & $+1$ & $\antifund$ & \cr
\hline
${\ov m}$, ${\ov \beta}$ & $-1$ & $\fund$ & \cr
\hline
${\tilde \eta}$ & $-1$ & & $\fund$\cr
\hline
${\ov{\tilde\eta}}$ & $+1$ & & $\antifund$ \cr
\hline
$Q$ & & $\fund$ & $\antifund$ \cr
\hline
${\tilde Q}$ & & $\antifund$ & $\fund$ \cr
\hline\hline
\end{tabular}
\end{center}}
where we have also included the 4d quark chiral multiplets $Q=q+\theta
\psi$, ${\tilde Q}={\tilde q}+\theta {\tilde \psi}$. Using results in
the literature e.g. \cite{Akerblom:2006hx}, we have the following fermionic couplings, which we already write in terms of 4d multiplets
\begin{equation}
{\ov \tau}\, {\ov \beta}\, m\, +\, {\ov \tau} \, \beta\, {\ov m}\, -\, 
\beta\, {\ov {\tilde Q}} \, {\tilde \eta} \, +\, {\ov \beta} \, {\ov Q} \, {\ov{\tilde \eta}} \, +\, {\ov m}\, {\ov{\tilde\eta}} \,{\ov D}{\ov Q} \, +\, m \, {\tilde \eta}\, {\ov{D{\tilde Q}}}
\label{longcoupling}
\end{equation}
Let us focus already on the case $N_f=N_c=1$. We can saturate all the
fermionic zero modes only by pulling down the first two and the last
two interactions in (\ref{longcoupling}). We obtain the analog of the Beasley-Witten result for our $N_f=N_c=1$ case
\begin{equation}
  |m|^4\, {\ov D}{\ov Q}\, {\ov {D{\tilde Q}}}
\end{equation}
Carrying out the bosonic zero mode integral, which introduces a
holomorphic function of the moduli and which we skip for simplicity,
the BPS instanton amplitude has the F-term structure
\begin{equation}
 \int\, d^2\theta\, e^{-T} \, {\ov D}{\ov Q} \, {\ov{
  D {\tilde Q}}}
\label{nfncbps}
\end{equation}

\subsubsection*{Coupling to the closed string sector}

Another important coupling for our considerations involving all of
moduli space is the coupling of the ${\ov \tau}$ zero modes coming from
the supersymmetrization of the instanton volume:
\begin{equation}
  S_{vol} = t + \theta \delta_\theta t + {\ov \tau} {\ov {D\Sigma}}
\end{equation}

We can use the last coupling in order to saturate some or all of the
${\ov \tau}$ zero modes. This gives rise to additional terms in the
low energy effective action compared to the Beasley-Witten expression
(\ref{nfncbps}), which comes just from the open string sector
couplings. For example:
\begin{equation}
  ({\ov \tau} {\ov \beta} m)\, ({\ov \tau} {\ov {D\Sigma}})\, (\beta\,
  {\ov {\tilde Q}} \, {\tilde \eta}) ({\ov m}\, {\ov{\tilde\eta}}
  \,{\ov D}{\ov Q}) \quad \longrightarrow \quad |m|^2 \, {\ov {\tilde Q}}\, (\ov {D\Sigma}\, \ov
  {DQ})
\end{equation}
Saturating the fermion zero modes in all possible ways, similar to the one above, and skipping the discussion of bosonic modes, we obtain a 4d instanton amplitude with the structure 
\beqa
  S_{4d} \, =  \, \int \, d^4x \, d^2\theta \, e^{-T}\,  \left(\,
 \ov {DQ} \, \ov {D\tilde Q}\,  +\, 
 \, {\ov Q}\, {\ov{D{\tilde Q}}}\,  \ov {D\Sigma} \, -
 \, {\ov {\tilde Q}}\, {\ov{DQ}}\, {\ov D\Sigma}\,
 -\,  {\ov Q}\,   {\ov {\tilde Q}}\,  \ov {D\Sigma}\, \ov {D\Sigma}\, 
  \right)\quad 
  \label{openclosedNfNc}
\eeqa
The meaning of the minus signs, and the F-term structure of this contribution will be clarified in next section.
This induced operator has an interesting form, in particular it couples the open and closed string sectors  in a nontrivial way. This structure can be clarified by deforming the configuration slightly away from the BPS regime, as we consider now.

\subsection{The near-BPS regime}
\label{nfncnear}

To address the behaviour of the system around the line  of BPS stability, and also to understand better the meaning of expression (\ref{openclosedNfNc}), let us consider the configuration slightly away from the BPS locus of the instanton. In this regime, the
misalignment can be described by the introduction of a FI term $\xi=\Sigma+{\ov\Sigma}$, both in the instanton and in the 4d gauge theory. In the latter, we can describe it as
\beqa
S_{FI}\, =\, \int\, d^4x\, d^4\theta\, (\,\Sigma\,+\,{\ov\Sigma}\,)\, V
\label{fi4d}
\eeqa
where we ignore a possible constant coefficient. As already mentioned, in the 4d gauge theory the resulting D-term can be canceled by 
either $Q$ or $\tilde Q$ acquiring a vev in order to make the abelian D-term potential vanish
\begin{equation}
  V_D\, = \,\left(\, |Q|^2\, -\, |\tilde Q|^2\, +\,\Sigma\,+\,{\ov\Sigma}\, \right)^2\, =\, 0
\end{equation}
Coming back to the stringy picture, the vev for (say) $Q$ tells us to
recombine the two slightly misaligned branes, and we end up in the
configuration depicted in Figure~\ref{nfnc}b.

The above discussion shows that from the string point of view, the
moduli space mixes the open and closed string sectors. One should
therefore find adapted coordinates which parameterize it
appropriately. As is familiar, the moduli space of D-flat directions,
modulo gauge transformations, can be parameterized using a set of gauge
invariant operators.  An important point in their construction is that
the field-dependent FI term (\ref{fi4d}) actually arises from a
modified Kahler potential for $\Sigma$ involving the vector multiplet
$V$, of the form (assuming canonical Kahler potential for simplicity)
\cite{Dine:1987xk}
\beqa
\int\, d^4x\, d^4\theta\, (\,\Sigma\,+\,{\ov\Sigma}\,-\, V)^2
\eeqa
Hence there is a non-trivial gauge transformation of $\Sigma$ by a shift of the gauge parameter. Namely, under $V\to V+\Lambda+{\ov \Lambda}$
\beqa
Q\,\to\, e^{i\Lambda}\,Q\quad ,\quad {\tilde Q}\,\to\,e^{-i\Lambda}\, {\tilde Q}\quad ,\quad \Sigma\to \Sigma-\Lambda
\eeqa
A suitable set of gauge invariants is then provided by
\beqa
B\,=\, Q\, e^\Sigma\quad ,\quad {\tilde B}\, =\, {\tilde Q}\, e^{-\Sigma}
\label{thebaryons}
\eeqa
These operators play the role of the baryons of the effective
``$SU(1)$'' theory below the scale of the $U(1)$ mass, generated by the
$B\wedge F$ coupling in (\ref{fi4d}). Notice for instance that
$B{\tilde B}=Q{\tilde Q}\equiv M$ is classically related to the
mesonic operator.

Hence, the discussion of the 4d instanton amplitudes is most naturally
carried out in terms of these variables. Of utmost importance for us
is that the structure of operators insertions in
(\ref{openclosedNfNc}) corresponds to the expansion of ${\ov{DB}}
\,{\ov{D{\tilde B}}}$. Thus, the F-term contribution of the instanton
at the BPS locus is given by
\beqa
S_{4d}  &=&  \int \, d^4x \, d^2\theta \, e^{-T}\, {\ov {DB}}\, {\ov{D{\tilde B}}}
\eeqa
This contribution is recovered as the ${\cal{O}}(\xi^0)$ term of the
instanton amplitude in the near-BPS regime, showing that the full
instanton amplitude is in a non-trivial class of the Beasley-Witten
cohomology. As in the example in Section \ref{isolated}, higher
order terms simply correspond to globally defined D-terms, and this
can be traced to the fact that the left over zero mode picks up a
component along ${\ov\tau}$, which itself picks up a component along
${\ov\theta}$.

\medskip

As we have argued, switching on a FI forces us to give a vev to $Q$ or
$\tilde Q$. From the point of view of the induced 4d operator,
switching on a FI corresponds to interpolating from a situation where
the first term in (\ref{openclosedNfNc}) dominates (the Beasley-Witten
gauge theory analysis) to a situation with a deeply non-BPS instanton
where the last term dominates. We study this regime of very non-BPS
instantons a bit more in the next section.

\subsubsection*{Vevs along the mesonic branch}
Before moving on to the non-BPS regime, let us shortly consider
another interesting direction in moduli space, namely that in which we
do not turn on a vev for $\Sigma$, and move instead along the
direction in which $Q$ and $\tilde Q$ get the same vev (i.e. the
mesonic branch in gauge theory language). In this situation the branes
recombine and move together along the plane of the black degeneration,
away from the instanton.

In this situation, the instanton and the branes get separated without
misaligning the system, and the $\ov\tau$ modes on the instanton no
longer get lifted. This is perfectly compatible with all of our
discussion so far. The setup is similar to the one in
Section~\ref{sec:U(1)-bps}, with the recombined brane playing the role
of the orientifold selecting the preferred $\NN=1$ subalgebra. From
the same considerations we expect the instanton to generate a ${\ov
  {D\Sigma}}\,{\ov{D\Sigma}}$ term, which is exactly the term that
dominates in (\ref{openclosedNfNc}) when both $Q$ and $\tilde Q$ get a
vev.

\subsection{The deep non-BPS  regime}

Let us now describe how the above structure of couplings reproduces
the expected physics in the deep non-BPS regime, namely for large
$\xi$. Here it is important that motion away from the BPS locus of the
instanton forces the recombination of the 4d spacefilling D-branes,
namely the scalar components in either $Q$ or ${\tilde Q}$, depending
on the sign of $\xi$, acquire a large vev, thereby removing a subset
of the instanton zero modes. Consider e.g. the regime where ${\tilde
  Q}$ acquires a large vev, and the modes $\beta$, ${\tilde \eta}$ are
removed (see eq. (\ref{longcoupling})). So are the bosonic modes $m$,
${\ov m}$ due to the misalignment of the instanton and the 4d
D-branes.  We are thus left with a $U(1)$ instanton, with the
universal sector of zero modes $\theta'$, ${\ov\tau}'$, and zero modes
${\ov \beta}$, ${\ov{\tilde \eta}}$ coupling to the 4d chiral
multiplet ${\ov Q}$ in the ``adjoint'' of the recombined $U(1)$, denoted $\Phi$ in what follows. In the
non-BPS regime, i.e. for non-zero BPS phase $\xi$, the universal set
can be traded for $\theta$, ${\ov\theta}$ to yield an instanton
amplitude roughly of the form
\begin{equation}
\int \, d^4x\, d^2\theta\, d^2{\ov\theta}\, e^{-T'} \, {\ov \Phi}\, +\, {\rm h.c.}
\label{nfncnbps}
\end{equation}
Here $T'$ is the expression of the wrapped volume as a
(non-holomorphic) function of the moduli. For instance, in the limit
where the instanton looks vertical in the picture of Figure
\ref{nfnc}, the instanton has tension but no (relevant) charge (in a
theory with orientifold, the orientifold image would cancel its charge
exactly), and $T'=\Sigma+{\ov \Sigma}$ where $\Sigma$ controls the
vertical position of the white degeneration.

The above is precisely the fermion zero mode content and amplitude
expected for a non-BPS $U(1)$ instanton wrapped on a cycle with a
chiral intersection with the 4d D-branes. It also matches
(\ref{openclosedNfNc}) in the limit where we discard the last two
terms since they are subleading when the instanton is very non-BPS. In
particular, saturating the ${\ov\theta}$'s with the antichiral pieces
in $T'$, one can generate amplitudes with two insertions of the 4d
fermions ${\ov\chi}$ in the $\ov {D\Sigma}$ multiplet, showing the
expected behaviour for a (non-isolated) $U(1)$ instanton.

\medskip

In conclusion, the instanton amplitude satisfies the overall global
picture discussed in Section \ref{isolated}, despite the fact that the
crucial modes ${\ov\tau}$ have non-trivial interactions. We expect a
similar discussion in other examples with interacting ${\ov\tau}$'s.

\section{Lifting of fermion zero modes and 4d supersymmetry breaking}
\label{lifting}

In this section we analyze the following possibility. It is in
principle possible to consider D-brane instantons which contribute to
the superpotential, and which nevertheless can misalign. This would
seem to contradict our general statements, based on the counting of
goldstinos. However, there is a way out, which automatically comes out
in a clever way in the explicit examples below. The modulus that takes
the instanton away from the BPS locus simultaneously triggers breaking
of 4d spacetime supersymmetry. Thus, the instanton actually does not
break any exact supersymmetry of the background, and is not forced to
have four goldstinos. In certain situations it still has two {\it
  approximate} fermion zero modes (which are not true goldstinos), and
the instanton generates 4d operators which, in the near supersymmetric
regime, can be thought of as an {\em approximate} superpotential in
the {\em approximately} supersymmetric theory, with susy breaking
broken by a spacetime D-term.

We discuss two examples. The first corresponds to a D-brane
realization of $N_f=N_c-1$ SQCD, where the instanton generates a
superpotential. Motion away from the BPS locus of the D-brane
instanton is parameterized by a closed string modulus, which
simultaneously induces a 4d spacetime D-term potential which cannot be
completely canceled and thus breaks 4d supersymmetry. The second
example corresponds to a $U(1)$ D-instanton with the additional
fermion zero modes lifted by closed string fluxes. Again, we find that
the modulus taking the instanton away from the BPS locus
simultaneously makes the closed string fluxes non-supersymmetric.

\subsection{Gauge theories with non-perturbative superpotential}
\label{suposusybreak}

\begin{figure}
\epsfysize=2.5cm
\begin{center}
\leavevmode
\epsffile{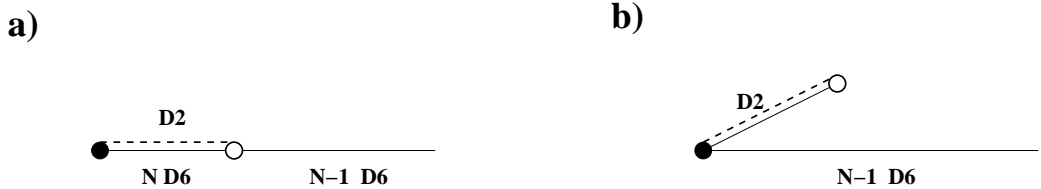}
\end{center}
\caption{\small D-brane realization of the $N_f=N_c-1$ SQCD theory in a geometry of the kind in Appendix \ref{oovafa}. The construction shows that instanton on its BPS locus (a) and away from it (b). The gauge D-branes cannot all recombine and define a non-supersymmetric background for the instanton.}
\label{susybreak}
\end{figure}

Let us consider a D-brane construction, similar to that of previous
section, of the $U(N_c)$ SQCD with $N_f=N_c-1$ flavours, see Figure
\ref{susybreak}a. In this case, the D-brane instanton on the BPS locus
generates the expected field theory Affleck-Dine-Seiberg
superpotential (see e.g. \cite{Akerblom:2006hx} for a detailed
computation in the string setup). Let us consider the effect of
misaligning the instanton.

In the regime of small misalignment, one can use the gauge theory viewpoint, where it is described as turning on a FI term. Being a D-term, this should not change the 4d superpotential. Thus it suggests that a non-BPS instanton induces a superpotential, seemingly in contradiction with our general picture, and with the counting of fermion zero modes from goldstinos. There is however no contradiction, as is manifest using the brane picture. In misaligning the instanton by moving the white degeneration up, as in Figure \ref{susybreak}b, the gauge D-branes cannot all recombine and one of the color brane misaligns (this is visible in the field theory description as an uncancelled 4d D-term). Thus away from the instanton BPS locus, there is really no supersymmetry in the background, and therefore no need for the instanton to have four goldstinos. Indeed, the modes ${\ov\tau}'$ are lifted by couplings with the bosonic zero modes between the instanton and the misaligned color D-brane. The instanton has only two (accidental) zero modes, the $\theta'$, and in the near BPS regime it induces operators which can be described in the field theory approximation as a superpotential (in a theory with D-term supersymmetry breaking).

The global description of this kind of example is very different from the previous systems, since in the deep non-BPS regime the background is very non-supersymmetric. We refrain from any further discussion.

\medskip

Another complementary approach to understand the $N_f=N_c-1$ SQCD system would be to start with the $N_f=N_c$ configuration, and add a mass term for one of the flavors. This was done by Beasley and Witten directly in field theory, the result being that the $N_f=N_c$ instanton is deformed so that the 4d operator it leads to has two less fermion insertions. In sketchy terms, we have (at the BPS locus)
\beqa
\int \, d^2\theta\, e^{-T}\, {\ov D}{\ov Q}\, {\ov D}{\ov {\tilde Q}} \, +\, \int \, d^2\theta\, M\, Q{\tilde Q} \, \longrightarrow \, \int \, d^2\theta\, e^{-T}
\label{bwsat}
\eeqa
This is in agreement with the fact that the IR of the $N_f=N_c$ theory with the mass term should reduce to the $N_f=N_c-1$ theory. On a similar line, we expect that when we turn on an FI deformation, there will be supersymmetry breaking in the $N_f=N_c$ theory with massive flavor. This is easy to see in field theory, and also in the brane realization of the theory.

\subsection{Closed string fluxes}

A well-known mechanism to lift zero modes is to consider D-brane instantons in the presence of $\cn=1$ closed string background fluxes \cite{Tripathy:2005hv,Bergshoeff:2005yp,Blumenhagen:2007bn}. As the fluxes break the Calabi-Yau $\cn =2$ supersymmetry down to $\cn=1$ on top of instanton worldvolume, there is a priori no reason for the additional zero modes $\bar{\tau}$ of Section \ref{isolated} to be present. Hence one would expect that, in the presence of fluxes, one can obtain $U(1)$ instantons that contribute to the non-perturbative superpotential. Since in principle such instantons can misalign, this would lead to contradiction with general arguments of counting of goldstinos. In the following we would like to argue, using the viewpoint of the world-volume fermionic action,  that this is not the case.

The Dirac action for the fermionic modes on a D$p$-instanton reads \cite{Martucci:2005rb}
\beq\label{faction}
\Theta\, P_{\k}^{Dp}\, \cd\, \Theta\ ,=\,
\Theta\, P_{\k}^{Dp}\left(\cam^{mn}\Gamma_m \cd_n-\frac12\co\right)\Theta
\eeq
where $\Theta$ are the fermionic superembedding variables, $P_{\k}^{Dp}$ is the $\kappa$-symmetry projector, which for an instanton with BPS phase $\xi$ selects the zero modes of the form (\ref{zerorot}), and $\cd_n$ and $\co$ are the operators appearing in the gravitino and dilatino variations, pull-backed to the D-instanton worldvolume. Finally, $\cam^{mn}$ is a matrix which depends on the instanton worldvolume flux $\cf$, and which for $\cf=0$ reduces to $g^{mn}$.

From (\ref{faction}) one can see that a instanton fermionic zero mode needs to satisfy two requirements. First, in order to be a true fermion mode (rather than a $\kappa$-symmetry parameter) it must not be projected out by $P_{\k}^{Dp}$; and, second, it must be annihilated by the operator $\cd$.
 Let us see how these requirements apply for the U(1) instanton of Section \ref{isolated}, which we will again assume to be rigid and isolated. Hence, the space of zero mode candidates is contained in Table \ref{zero}. It is important that, since the instanton is not mapped to itself under the orientifold, one can use (\ref{faction}) on the covering space.

If our background is Calabi-Yau, it is clear that any linear combination of the zero modes in table \ref{zero} is annihilated by $\cd$, for they are all individually annihilated by $\cd_n$ and $\co$. Hence, the rotated modes (\ref{zerorot}) selected by $P_{\k}^{Dp}$ are automatically zero modes, whether we are in the BPS locus $(\xi = 0 \,{\rm or\,} \pi)$ or away from it, and we recover the well-known result that such $U(1)$ instanton has four zero modes.

If we now turn on background fluxes the operators $\cd_n$ and $\co$ get deformed, and it is no longer true that $\cd_n{\tau} = \cd_n\bar{\tau} = \co {\tau} = \co \bar{\tau} = 0$. 
However, we can use the fact that for $\xi \neq 0$, the instanton is non-BPS and thus has four zero modes, which are the four goldstinos of the broken $\cn =1$ supersymmetry, namely ${\cal D}\bar\tau'=0$ and ${\cal D}\th'=0$, to yield
\beq
0=\cd {\ov \tau}' \, = \, \cos (\xi/2) \, \cd {\ov\tau}\, +\, \sin(\xi/2)\,\cd {\ov \theta} \, =\, \cos (\xi/2) \, \cd {\ov\tau}\, = 0,
\label{almendruco}
\eeq
and similarly for $\th'$. Here we have used that $\xi$ is constant and that $\cd_n {\ov \th} = \co {\ov \th} = 0$, for this mode comes from the background Killing spinor. 

We thus find that, anywhere close to the BPS locus $\xi = 0$, ${\ov \tau}$ is annihilated by $\cd$. By continuity of the spectrum of such differential operator, we conclude that it must also be so for $\xi = 0$. Thus, it is not possible to turn on this kind of closed string backgrounds to lift the additional fermion zero modes ${\ov\tau}$.

Notice that the argument above does not apply if the instanton cannot misalign via a $\xi \neq 0$. This includes the case of instantons invariant under some orientifold action, of gauge group either $O(1)$ or $USp(2)$, that were considered in \cite{Bergshoeff:2005yp} to find new contributions to the superpotential.\footnote{In addition, as pointed out in \cite{Bergshoeff:2005yp}, in this case the fermionic action (\ref{faction}) is not valid, and needs to be further projected by the orientifold action.}

Notice that there is a further way out of the argument, involving breaking of 4d supersymmetry in a way similar to section \ref{suposusybreak}. There may exist flux background which lift the additional fermion zero modes, and allow the instanton to contribute to the superpotential, if the flux background is non-supersymmetric for non-zero $\xi$. Indeed, in that case taking $\xi \neq 0$ would not only mean that the instanton is non-BPS, but also that the closed string background is $\cn = 0$. In that case (\ref{almendruco}) need not be true because, away from $\xi = 0$, $\cd_n {\ov \th} = \co  {\ov \th} = 0$ no longer holds. In addition, we cannot claim that there is a minimum of four fermion zero modes coming from goldstinos, for there are no bulk supersymmetries to be broken.

To our knowledge, lifting of fermion zero modes by the latter kind of fluxes has not been much considered in the literature. As a sketchy example, along the lines of section 5.3 of \cite{Blumenhagen:2007bn}, we may consider a toroidal orientifold compactification\footnote{For simplicity we ignore tadpole cancellation, or include suitable antibranes away from the system of interest so that they do not modify the argument.} with $J=\sum dz_i\, d{\ov z}_i$ and a supersymmetric primitive $(2,1)$ 3-form flux $G_{{\bar 1}23}$. 
Consider a D3-brane instanton wrapped on $z_1,z_2$ and magnetized with a world-volume primitive flux $F_{1{\bar 2}}$, so that it is BPS. As argued in \cite{Blumenhagen:2007bn} this flux could remove the additional fermion zero modes in the universal sector ${\ov\tau}$. This would seem to contradict our general arguments, since the BPS phase of the D3-brane can be misaligned by turning on a component of the Kahler form along $d{\bar z}_1\,dz_2\, +\, dz_1\, d{\bar z}_2$, so that the world-volume flux is no longer primitive and the instanton becomes non-BPS. However, this simultaneously makes the 3-form flux non-primitive, and thus breaks spacetime supersymmetry. 

For the most studied setup of type IIB with 3-form fluxes on warped genuine Calabi-Yau geometries \cite{Dasgupta:1999ss,Giddings:2001yu}, the 3-form flux primitivity condition is automatic and the fluxes do not affect the (K\"ahler) moduli controlling the FI parameter $\xi$.  Hence, by our discussion above, they will never succeed in lifting the zero modes ${\ov\tau}$ unless they break supersymmetry at the same time. This was partially observed for IIB D3-brane instantons in 3-form flux backgrounds in \cite{Blumenhagen:2007bn}, but should hold in full generality. Note that the above analysis provides a new, deeper understanding  of the negative results in \cite{Tripathy:2005hv,Bergshoeff:2005yp,Blumenhagen:2007bn}.

\subsection{Effects of additional instantons}

A further possibility is to consider the lifting or absorption of additional fermion zero modes of an instanton by another. Refraining from a general analysis, let us briefly sketch an example of the interplay between lifting of additional zero modes and BPS misalignment of  $U(1)$ instanton.

In \cite{GarciaEtxebarria:2007zv} a mechanism to saturate the extra fermion zero modes of a $U(1)$ instanton was presented. Namely, a $U(1)$ instanton in the presence of an additional $O(1)$ instanton can produce a two-instanton effect which contributes to the superpotential. In a precise sense, the $O(1)$ instanton produces a non-perturbative lifting of the additional zero modes of the $U(1)$ instanton. This would seem to contradict our general argument above, since there is a parameter which can misalign the $U(1)$ instanton and make it non-BPS. The resolution of the puzzle is that the misalignment is realized when only the $U(1)$ instanton is present; in the presence of the $O(1)$ instanton, namely precisely when the additional zero modes are lifted, the same parameter actually turns the two-instanton system into a BPS one-instanton system. Namely, there is line of marginal stability for the $U(1)$ instanton (with four fermion zero modes), which is just a line of threshold stability for the $U(1)$-$O(1)$ two-instanton system (with two fermion zero modes). Thus no contradiction with our general picture is found. 

This provides an amusing example of a behaviour complementary to the previous systems. In this case, the mechanism that provides the lifting of extra zero modes does not break the 4d supersymmetry, but rather restores the BPS property for the new multi-instanton system.

It is easy to device other systems of several instantons and show that such systems manage in clever ways to always comply as expected with the general rules of counting of goldstinos. Other examples following a similar pattern have appeared in \cite{Cvetic:2008ws}.

\section{Conclusions}
\label{conclu}

In this paper we have completed the picture of instanton amplitudes as
they cross lines of BPS stability of different kinds.  BPS instantons
contributing to the superpotential have at worst lines of threshold
stability, where they split into mutually BPS instantons, which
reconstruct the non-perturbative superpotential. BPS instantons
contributing to higher F-terms can also have lines of marginal
stability, beyond which they turn into (possibly multi-instanton)
non-BPS systems, which reproduce the same F-term (rewritten locally as
a D-term), modulo globally defined D-terms.

The picture is consistent, in a non-trivial way, with standard wisdom
of instanton fermion zero mode counting, and with holomorphy of 4d
$\NN=1$ F-terms.  An important lesson in this story has been the role
of the Beasley-Witten cohomological structure of higher
F-terms. Although the structure of this cohomology is unfamiliar and
quantitative statements are hard to make explicit, the instanton zero
mode structure automatically reproduces the appropriate features. It
would be interesting to develop a formalization of these results,
beyond the examples we have provided.

It would also be interesting to continue understanding the behaviour
of instanton amplitudes globally in moduli space. In $\NN=2$ language
we have focused on hypermultiplet moduli space (complex structure
moduli space for IIA, Kahler for IIB). It would be interesting to also
explore further the dependence on vector multiplet moduli space
(Kahler for IIA, complex for IIB), which essentially control the
1-loop prefactors of the exponential term in the instanton
amplitude. These moduli are naturally related to F-terms on the
world-volume of the instanton, and therefore the holomorphic
dependence on the moduli is quite straightforward. Still one may
expect interesting lessons also from a deeper look into this
dependence, and we hope that some of the concepts we discussed in this
work will be useful in this new setup.

For instance, it is easy to consider systems of BPS systems which can
become non-BPS due to an uncancelled world-volume F-term. Possibly the
simplest system of this kind is given by a D1-brane instanton wrapped
on the non-trivial two-cycle of a two-center hyperkahler ALE
geometry. There is a triplet of blow-up parameters, which, with
respect to some preferred 4d $\NN=1$ supersymmetry, couple to the
instanton as a real world-volume FI term, and a holomorphic
world-volume F-term (a superpotential linear in a complex bosonic zero
mode). As in our examples of D-term misalignment in this paper, the
F-term misaligned instanton has four goldstinos and generates a 4d
non-perturbative operator, which is writable as a D-term locally in
moduli space and which reduces to a 4d non-perturbative higher F-term
as one moves the BPS locus. In this case, the picture of F-term
misalignment is identical to D-term misalignment since the underlying
hyperkahler geometry implies a tri-holomorphic symmetry relating the
different components in the triplet of blowing-up parameters. We
nevertheless expect a similar behaviour for F-term misalignment in
more generic situations, since the basic facts relate to general
properties of counting of goldstinos and properties of higher F-terms.

We expect interesting forthcoming fundamental results and interesting applications from  continuing the analysis of instanton amplitudes globally in moduli space.

{\bf Acknowledgements}\\
We thank L. Ib\'a\~nez, T. Weigand for useful discussions. A.M.U. thanks M. Gonz\'alez for encouragement and support. I.G.-E. thanks the CERN Theory division for
hospitality, and N. Hasegawa for kind support. This work has been
supported by the European Commission under RTN European Programs
MRTN-CT-2004-503369, MRTN-CT-2004-005105, by the CICYT (Spain) and the
Comunidad de Madrid under project HEPHACOS P-ESP-00346. The work of
I.G.-E. was financed by the Gobierno Vasco PhD fellowship program.

\appendix

\section{Some useful geometries}
\label{oovafa}

Here we describe a set of geometries, introduced in
\cite{Ooguri:1997ih} and already used in
\cite{GarciaEtxebarria:2007zv} in a similar context, and which we use
in several of our explicit examples. They are non-compact geometries,
but they suffice to provide instanton effects and transitions as long
as they involve just the local structure of compact cycles.

Consider the class of local Calabi-Yau manifolds, described by
\begin{eqnarray}
xy= \prod_{k=1}^P (z-a_k) \nonumber \\
x'y'= \prod_{k'=1}^{P'} (z-b_k')
\end{eqnarray}
It describes two $\IC^*$ fibrations, parameterized by $x,y$ and $x',y'$, varying over the complex plane $z$, and degenerating at the locations $a_i$, $b_i$ respectively. The local geometry contains lagrangian 3-cycles obtained by fibering the two
$\IS^1$'s in the two $\IC^*$ fibers over segments joining pairs of degeneration points on the base. Segments joining pairs of $a$-type degenerations or pairs of $b$-type degenerations lead to 3-cycles with topology $\IS^2\times \IS^1$, while segments joining $a$- and $b$-type degenerations lead to 3-cycles with topology $\IS^3$. We denote $[p_1,p_2]$ the 3-cycle associated to the pair of degeneration points $p_1$, $p_2$.

Introducing the holomorphic 3-form 
\begin{eqnarray}
\Omega \, = \, \frac{dx}{x} \, \frac{dx'}{x'}\, dz
\end{eqnarray}
the 3-cycle $[p_1,p_2]$ is calibrated by the form $e^{i\theta}\Omega$,
where $\theta$ is the angle of the segment $[p_1,p_2]$ with the real
axis in the $z$-plane. Namely $Im(e^{i\theta} \Omega)|_{[p_1,p_2]}=0$,
where $|_{[p_1,p_2]}$ denotes restriction to the 3-cycle. Segments parallel in the $z$-plane define 3-cycles preserving a common supersymmetry. Our configurations will be d $\NN=1$ supersymmetric, with the preferred supersymmetry associated to segments parallel to the real axis in $z$.

We will consider stacks of 4d spacefilling D6-branes and/or euclidean D2-branes wrapping the different 3-cycles, and describe the non-perturbative superpotentials arising from these configurations. The open string modes and their interactions are easy to determine. For instance, each stack of $N$ D6-branes on a 3-cycle leads to a $U(N)$ gauge group in a vector multiplet of ${\cal N}=1$ supersymmetry for 3-cycles of
$\IS^3$ topology, and of ${\cal N}=2$ supersymmetry for 3-cycles of
$\IS^2\times \IS^1$ topology. The angle $\theta$ introduced above
determines the precise supersymmetry preserved by the corresponding
set of branes. Also, two D6-branes wrapping two 3-cycles involving one
common degeneration point lead to a vector-like pair of bi-fundamental
chiral multiplets, arising from open strings in the intersection of
3-cycles (which is topologically $\IS^1$, coming from the $\IC^*$ that
does not degenerate at the intersection).

As discussed in \cite{Ooguri:1997ih} one can perform T-dualities along the two $\IS^1$ directions, and map the configuration to a Hanany-Witten setup of $P$ NS-branes (along 012345) and $P'$ NS'-branes (along 012389), with D4-branes (along 01236) suspended among them, in a flat space geometry with a periodic coordinate $x^6$. The gauge theory content described above follows from the standard rules in this setup (see \cite{Giveon:1998sr}). This picture also facilitates the computation of the superpotential, whose general discussion we skip, but which we present in our concrete example below.

\end{document}